\documentclass[twocolumn,twocolappendix]{aastex7}

\pdfoutput=1
\usepackage{amsmath,amstext}
\usepackage[T1]{fontenc}
\usepackage{times}
\usepackage[figure,figure*]{hypcap}
\usepackage{natbib}
\usepackage{hyperref}
\usepackage{xspace}
\usepackage{booktabs}
\usepackage{graphicx}
\usepackage{multirow}
\usepackage[encapsulated]{CJK}
\usepackage[caption=false]{subfig}
\graphicspath{{./}{Figures/}}
\newcommand{\grizli}{\textsc{Gri$z$li}\xspace}
\newcommand{\eazy}{\textsc{EA$z$Y}\xspace}
\newcommand{\pros}{\textsc{Prospector}\xspace}
\newcommand{\dynesty}{\textsc{dynesty}\xspace}
\newcommand{\prosb}{\pros{}-$\beta$\xspace}

\newcommand{\lgmstar}{log(M$_{\star}$/M$_{\odot}$)}
\newcommand{\pysersic}{\textsc{pysersic}\xspace}

\received{XXX}
\revised{XXX}
\accepted{XXX}
\submitjournal{the Astrophysical Journal Letters}
\shorttitle{No Evidence of Environmental Quenching at z$\sim$2.6}
\shortauthors{Pan et al.}

\begin{document}
\begin{CJK*}{UTF8}{gbsn}
\title{UNCOVER/MegaScience: No Evidence of Environmental Quenching in a z$\sim$2.6 Proto-cluster}

\correspondingauthor{Richard Pan}
\email{richard.pan@tufts.edu}
\author[0000-0002-9651-5716]{Richard Pan}
\affiliation{Department of Physics \& Astronomy, Tufts University, Medford, MA 02155, USA}
\email{richard.pan@tufts.edu}

\author[0000-0002-1714-1905]{Katherine A. Suess}
\affiliation{Department for Astrophysical and Planetary Science, University of Colorado, Boulder, CO 80309, USA}
\email{wren.suess@colorado.edu}

\author[0000-0001-9002-3502]{Danilo Marchesini}
\affiliation{Department of Physics \& Astronomy, Tufts University, Medford, MA 02155, USA}
\email{danilo.marchesini@tufts.edu}

\author[0000-0001-9269-5046]{Bingjie Wang (王冰洁 
 \ignorespacesafterend)}
\affiliation{Department of Astronomy \& Astrophysics, The Pennsylvania State University, University Park, PA 16802, USA}
\affiliation{Institute for Computational \& Data Sciences, The Pennsylvania State University, University Park, PA 16802, USA}
\affiliation{Institute for Gravitation and the Cosmos, The Pennsylvania State University, University Park, PA 16802, USA}
\email{bwang@psu.edu}

\author[0000-0001-6755-1315]{Joel Leja}
\affiliation{Department of Astronomy \& Astrophysics, The Pennsylvania State University, University Park, PA 16802, USA}
\affiliation{Institute for Computational \& Data Sciences, The Pennsylvania State University, University Park, PA 16802, USA}
\affiliation{Institute for Gravitation and the Cosmos, The Pennsylvania State University, University Park, PA 16802, USA}
\email{joel.leja@psu.edu}

\author[0000-0002-7031-2865]{Sam E. Cutler}
\affiliation{Department of Astronomy, University of Massachusetts, Amherst, MA 01003, USA}
\email{secutler@umass.edu}

\author[0000-0001-7160-3632]{Katherine E. Whitaker}
\affiliation{Department of Astronomy, University of Massachusetts, Amherst, MA 01003, USA}
\affiliation{Cosmic Dawn Center (DAWN), Niels Bohr Institute, University of Copenhagen, Jagtvej 128, K{\o}benhavn N, DK-2200, Denmark}
\email{kwhitaker@astro.umass.edu}

\author[0000-0001-5063-8254]{Rachel Bezanson}
\affiliation{Department of Physics and Astronomy and PITT PACC, University of Pittsburgh, Pittsburgh, PA 15260, USA}
\email{rachel.bezanson@pitt.edu}

\author[0000-0002-0108-4176]{Sedona H. Price}
\affiliation{Department of Physics and Astronomy and PITT PACC, University of Pittsburgh, Pittsburgh, PA 15260, USA}
\email{sedona.price@pitt.edu}

\author[0000-0001-6278-032X]{Lukas J. Furtak}
\affiliation{Physics Department, Ben-Gurion University of the Negev, P.O. Box 653, Be’er-Sheva 84105, Israel}
\email{furtak@post.bgu.ac.il}

\author[0000-0003-1614-196X]{John R. Weaver}
\affiliation{Department of Astronomy, University of Massachusetts, Amherst, MA 01003, USA}
\email{john.weaver.astro@gmail.com}

\author[0000-0002-2057-5376]{Ivo Labb\'e}
\affiliation{Centre for Astrophysics and Supercomputing, Swinburne University of Technology, Melbourne, VIC 3122, Australia}
\email{ilabbe@swin.edu.au}

\author[0000-0003-2680-005X]{Gabriel Brammer}
\affiliation{Cosmic Dawn Center (DAWN), Niels Bohr Institute, University of Copenhagen, Jagtvej 128, K{\o}benhavn N, DK-2200, Denmark}
\email{gabriel.brammer@nbi.ku.dk}

\author[0000-0001-6454-1699]{Yunchong Zhang} 
\affiliation{Department of Physics and Astronomy and PITT PACC, University of Pittsburgh, Pittsburgh, PA 15260, USA}
\email{yunchongzhang@pitt.edu}

\author[0000-0001-8460-1564]{Pratika Dayal}
\affiliation{Kapteyn Astronomical Institute, University of Groningen, 9700 AV Groningen, The Netherlands}
\email{p.dayal@rug.nl}

\author[0000-0002-1109-1919]{Robert Feldmann}
\affiliation{Department of Astrophysics, University of Zurich, Zurich CH-8057, Switzerland}
\email{robert.feldmann@uzh.ch}

\author[0000-0002-5612-3427]{Jenny E. Greene}
\affiliation{Department of Astrophysical Sciences, Princeton University, 4 Ivy Ln., Princeton, NJ 08544, USA}
\email{jgreene@astro.princeton.edu}

\author[0000-0002-3254-9044]{Karl Glazebrook}
\affiliation{Centre for Astrophysics and Supercomputing, Swinburne University of Technology, PO Box 218, Hawthorn, VIC 3122, Australia}
\email{kglazebrook@swin.edu.au}

\author[0000-0001-8367-6265]{Tim B. Miller}
\affiliation{Center for Interdisciplinary Exploration and Research in Astrophysics (CIERA), Northwestern University, 1800 Sherman Ave, Evanston IL 60201, USA}
\email{timothy.miller@northwestern.edu}

\author[0000-0001-7300-9450]{Ikki Mitsuhashi}
\affiliation{Department for Astrophysical and Planetary Science, University of Colorado, Boulder, CO 80309, USA}
\email{ikki.mitsuhashi@colorado.edu}

\author[0000-0002-9330-9108]{Adam Muzzin}
\affiliation{Department of Physics and Astronomy, York University, 4700 Keele Street, Toronto, Ontario, ON MJ3 1P3, Canada}
\email{muzzin@yorku.ca}

\author[0000-0003-2804-0648 ]{Themiya Nanayakkara}
\affiliation{Centre for Astrophysics and Supercomputing, Swinburne University of Technology, Melbourne, VIC 3122, Australia}
\email{themiyananayakkara@gmail.com}

\author[0000-0002-7524-374X]{Erica J. Nelson}
\affiliation{Department for Astrophysical and Planetary Science, University of Colorado, Boulder, CO 80309, USA}
\email{erica.nelson-1@colorado.edu}

\author[0000-0003-4075-7393]{David J. Setton}
\thanks{Brinson Prize Fellow}
\affiliation{Department of Astrophysical Sciences, Princeton University, 4 Ivy Ln., Princeton, NJ 08544, USA}
\email{davidsetton@princeton.edu}

\author[0000-0002-0350-4488]{Adi Zitrin}
\affiliation{Physics Department, Ben-Gurion University of the Negev, P.O. Box 653, Be’er-Sheva 84105, Israel}
\email{zitrin@bgu.ac.il}

\begin{abstract}

Environmental quenching -- where interactions with other galaxies and/or the intra-cluster medium (ICM) suppress star formation in low-mass galaxies -- has been well-established as the primary driver behind the formation of the red sequence for low-mass galaxies within clusters at low redshift ($z<1$). However, it remains unclear whether these mechanisms are active at higher-redshifts in proto-cluster environments that are not yet fully virialized.” In large part, this regime has remained unexplored due to observational limitations; however, JWST has recently opened a new window into the role of environmental quenching on low-mass (\lgmstar$<$9.0) galaxies at cosmic noon ($2 < z < 3$). Here, we leverage the deep imaging and R~$\sim$ 15 spectrophotometry enabled by the 20 band JWST/NIRCam data from the UNCOVER and MegaScience programs to examine environmental quenching in a newly discovered $z\approx2.58$ proto-cluster. We compare the star formation histories (SFHs) of 19 low-mass quiescent galaxies in the proto-cluster to a matched sample of 18 in the field, and find no significant differences. This similarity extends to galaxy sizes and quenched fractions, which also show no significant differences between the two environments across the full stellar mass range (8.5$<$\lgmstar$\leq$11.0). This indicates that the proto-cluster has not yet accelerated quenching relative to the field and is consistent with expectations that $z>2$ proto-clusters have yet to virialize and develop a dense enough environment required to efficiently quench low-mass galaxies.

\end{abstract}



\section{Introduction}\label{sec:intro}

Nothing lasts forever. For many galaxies, their star formation shuts down -- a process often referred to as ``quenching''. The physical processes responsible for quenching, and how they operate in galaxies across different stellar masses and redshifts, are not yet fully understood \citep{manbelli2018}. The proposed quenching mechanisms are roughly separated into internal and external processes \citep{peng2010b}. Internal processes include quenching via stellar winds, supernovae feedback, and AGN feedback \citep{Ciotti1991, Croton2006, Ishibashi2012}. External processes, commonly known as environmental quenching, include quenching via ram-pressure stripping, galaxy interactions, starvation, and harassment, and primarily affect low stellar mass galaxies \citep{gunngott1972, farouki1981,moore1996, Balogh2004, peng2015, albertsnoble2022}. 

Environmental quenching can be driven by the buildup of a hot intra-cluster medium (ICM) \citep{sarazin1986}. The ICM suppresses star formation in in-falling galaxies through two main mechanisms. On short timescales, ram-pressure stripping can remove cold gas from galaxies as they move through the dense ICM \citep{crowl2008, muzzin2014, boselli2014, steinhauser2016,cortese2021, boselli2022}. On longer timescales, the hot ICM can heat the diffuse gas in the dark matter halo, preventing the accretion of more cold gas and leading to quenching via starvation \citep{peng2015, albertsnoble2022}. Environmental quenching via galaxy interactions or harassment can also occur when galaxy fly-bys or mergers induce instabilities in the disks that trigger central starbursts and AGN activity to rapidly deplete the cold gas reservoirs \citep{smith2010,bialas2015, rodriguezmontero2019}.

These environmental processes have been observed to efficiently quench galaxies with stellar masses of \lgmstar$<$9.5, thereby building-up the red sequence in lower-redshift ($z\leq 1.5$) clusters \citep{Dressler1980, Balogh2004, Kauffmann2004, Baldry2006, peng2010b}. A common approach to inferring the impact of environmental quenching is to compare the ``quenched fraction'' -- the fraction of all galaxies that are quiescent for a given stellar mass range -- between different environments \citep{peng2010b,Kawinwanichakij2016,kawinwanichakij2017,tomczak2017,vanderburg2020}. The assumption is that low-mass quiescent galaxies in dense galaxy clusters would be still forming stars if they lived in a less dense environment. Instead their star formation is truncated early due to these environmental processes, resulting in systematically older stellar populations compared to similar-mass field galaxies. This contributes to both the observed age discrepancy and the elevated quenched fraction in overdense environments \citep{trager2000,vandokkum2003, thomas2005, renzini2006, thomas2010, muzzin2014,  paulinoafonso2020, mcnab2021}.

While these mechanisms operate in galaxy clusters at lower redshift, it is unclear whether they have a significant impact at cosmic noon. These cosmic structures undergo a major transition between $1.5 < z \lesssim 2.5$, going from dynamically evolving, unvirialized proto-cluster structures towards virialized clusters in hydrostatic equilibrium \citep{muldrew2015,wang2016,chiang2017, champagne2021}. During this transitional epoch, proto-clusters are forming the bulk of their total stellar mass ($\sim65\%$), undergoing virialization leading to stable equilibrium, and assembling smaller dark-matter halos onto the central halo as they develop their hot ICM \citep{chiang2013, muldrew2015, overzier2016,chiang2017}. To date, most studies of proto-cluster galaxy populations at cosmic noon have only been able to examine galaxies with stellar masses of \lgmstar$>$9.5, due to difficulty with observing faint low-mass galaxies at large cosmic distances, especially quiescent ones \citep{Kawinwanichakij2016, kawinwanichakij2017,tomczak2017, vanderburg2020, mcconachie2022, edward2024, forrest2024, singh2024, naufal2024}. Even at these high-masses there is still disagreement on the role of environmental quenching as some studies find that the quenched fractions are consistent with those found in the field \citep{edward2024, forrest2024} while other studies see a significantly higher quenched fraction especially at ultra high-masses \citep[i.e. \lgmstar$>$11.0;][]{mcconachie2022, ito2023, naufal2024}.

Until JWST, the low-mass quiescent population at cosmic noon was below the observational detection limit of even HST \citep{tal2014}, in part due to the lack of deep imaging needed to detect these faint galaxies. These limitations were especially pronounced for low-mass quiescent galaxies in proto-clusters, where environmental quenching is expected to be most effective \citep{albertsnoble2022}. Since then, JWST has dramatically opened up new windows into the population of intermediate- to low-mass quiescent galaxies at cosmic noon, both through photometric \citep{cutler2024, hamadouche2024, alberts2024} and spectroscopic \citep{marchesini2023, sato2024} studies. 

So the question is: can proto-cluster environments efficiently quench low-mass galaxies even before they undergo this big transition, or can only fully virialized massive clusters drive environmental quenching of low-mass galaxies? We can infer the role of environmental quenching by examining potential differences in the star-formation histories of quiescent galaxies in an overdense environment versus the field.

In this Letter, we present a serendipitously discovered $z\approx2.58$ proto-cluster behind the Abell 2744 (A2744) lensing field. We use data from the JWST Ultradeep NIRSpec and NIRCam ObserVations before the Epoch of Reionization (UNCOVER) Cycle 1 Treasury Program \citep{bezanson2024} and the Medium Bands, Mega Science Cycle 2 survey \citep{suess2024}. We leverage the unprecedented depth ($\sim$29.5 AB to $\sim$32.0 AB with magnification; \citealt{weaver2024a}) and the constraining power of the NIRCam medium band photometry ($\sigma_{\text{NMAD}} \sim 0.015$; \citealt{suess2024}) to robustly infer the redshifts and stellar populations \citep{wang2024}. This allows us to identify proto-cluster candidates and their low-mass quiescent members. In total, we identify 217 members (34 quiescent) in this $z\approx2.58$ proto-cluster. We compare their SFHs, formation/quenching times, sizes, and quenched fractions to those of their field counterparts (216 members and 29 quiescent galaxies) and explore, for the first time, the effects of the environment on low-mass quiescent galaxy evolution at cosmic noon. The Letter is structured as follows: in Section \ref{sec:methods} we present the proto-cluster and its quiescent members, in Section \ref{sec:analysis} we detail the population modeling done to statistically compare the timescales of the two populations, and in Section \ref{sec:discussion} we discuss the findings in relation to current works. In this study, we adopt the best-fit cosmological parameters from the WMAP9 cosmology \citep{Hinshaw2013} with $H_0=69.32~{\rm km~s^{-1}~Mpc^{-1}}$, $\Omega_M=0.2865$, and $\Omega_\Lambda=0.7135$, we present all magnitudes in the AB system, and we utilize a Chabrier \citep{chabrier2003} initial mass function.

\section{Methods} \label{sec:methods}
\begin{figure*}
    \centering
    \includegraphics[width=\linewidth]{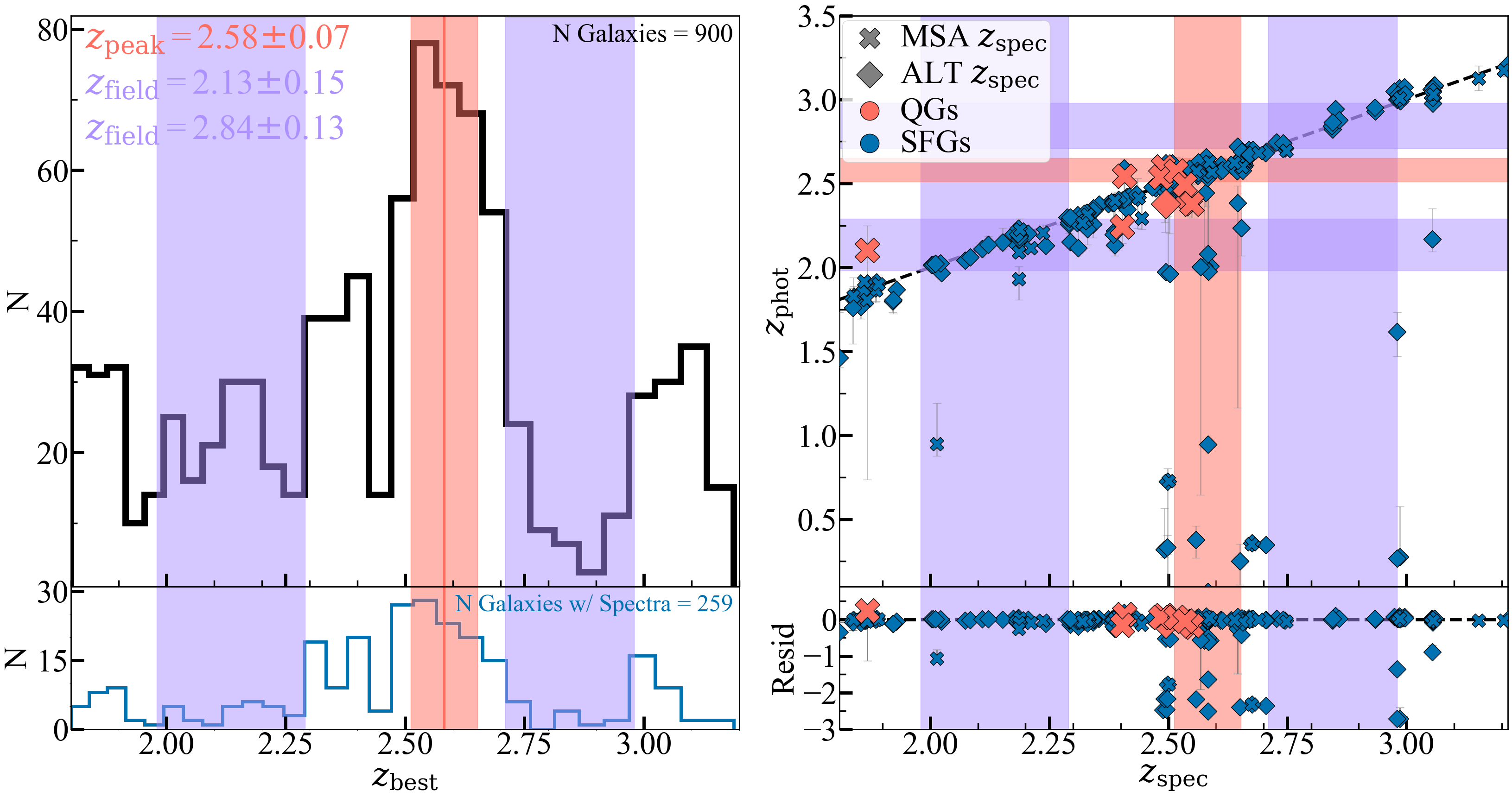}
    \caption{Top left: Redshift histogram of all galaxies in the redshift range centered around cosmic noon adopting the ``best'' available redshift as detailed in section \ref{subsec:data}. Bottom left: Redshift histogram from re-fitting the photometry with a uniform redshift prior at $z_{\text{spec}}\pm{0.1}$. The overdensity is centered at $z\approx2.58$ (red) while the field was selected at higher and lower redshifts. Top right: Comparison of the photo-$z$'s and spec-$z$'s of 259 spectroscopically observed galaxies. The quiescent galaxies are colored in red, the star-forming galaxies in blue, and the UNCOVER objects shown as crosses and the ALT objects as diamonds. Bottom right: Residuals of the spectroscopically observed galaxies ($z_{\text{phot}} - z_{\text{spec}}$). The red and purple stripes denote the overdensity and field redshift ranges. Our $\sigma_{\text{NMAD}}\approx0.010$ is consistent with a $\sigma_{\text{NMAD}}\approx0.015$ found in \citet{suess2024} and highlights the constraining power of the medium band photometry.}
    \label{fig:z_hist}
\end{figure*}

\subsection{Data}\label{subsec:data}
We use the DR3\footnote{The photometric catalog is available on our website (\url{https://jwst-uncover.github.io/DR3.html}) and Zenodo \citep{photo2024}.} photometric catalog from the UNCOVER (JWST-GO-2561, PIs: Bezanson \& Labb\'e; \citealt{bezanson2024}) and MegaScience (JWST-GO-4111, PI: Suess; \citealt{suess2024}) surveys. The dataset contains all 20 NIRCam broad- and medium-band filters, including data from GLASS-ERS \citep{treu2022}, ALT (PIs: Matthee \& Naidu; \citealt{naidu2024}), MAGNIF (PI: Sun; \citealt{li2023}), GO-3538 (PI: Iani), and GO-2756 \citep[PI: Chen;][]{chen2024}. We also utilize HST data in F435W, F606W, F814W, F105W, F140W, and F160W from the HFF (PI: Lotz; \citealt{lotz2017}) and BUFFALO (PI: Steinhardt; \citealt{steinhardt2020}) programs. We use the ``super'' aperture photometric catalog, which matches the best aperture to the isophotal area of each galaxy, as described in detail by \citealt{weaver2024a}. 

We fit the photometry to derive galaxy properties using the \prosb model from \citet{wang2023}, which uses the \pros Bayesian inference framework \citep{johnson2021}. The \prosb model robustly infers the redshift, rest-frame colors, and stellar population properties such as stellar masses and SFHs \citep{wang2023}. Compared to \pros-$\alpha$ \citep{leja2017}, the \prosb model \citep{wang2023} includes  three
new priors on the stellar mass, number density, and SFH. The first is a stellar mass function prior based on the \cite{leja2020} stellar mass functions. The second is a galaxy number density prior based on a mock catalog of galaxies from the JAGUAR simulations \citep{williams2018}. The last is a SFH prior that matches the expectation value in each SFH bin (lookback time) to the cosmic star formation rate densities given from \cite{behroozi2019}. Including these priors better constrains galaxy properties across the broad parameter space probed by deep JWST photometric surveys \citep{wang2023}. The \pros framework uses the \dynesty dynamic nested sampling package \citep{speagle2020} to sample draws from the prior distribution \citep{johnson2021}. We adopt \texttt{v2.0} of the parametric strong lensing model of A2744 by \citet{furtak2023}, which was recently updated with UNCOVER spectroscopic redshifts in \citet{price2024}, to compute magnification factors and correct the flux when fitting for the redshift and stellar population properties. 

When available we include spectroscopic redshifts ($z_{\text{spec}}$) in place of photometric redshifts ($z_{\text{phot}}$). The spec-$z$'s are provided by the UNCOVER DR4\footnote{The spectroscopic catalog is available on our website (\url{https://jwst-uncover.github.io/DR4.html}) and Zenodo \citep{spec2024}.} spectroscopic redshift catalog \citep{price2024} and the ALT DR1 redshift catalog \citep{naidu2024}. The UNCOVER NIRSpec/MSA follow-up survey targeted $\sim$700 objects in A2744 and the reduction was done using \textsc{msaexp} \citep{msaexp2022}. Of these, we utilize $\sim$400 objects which have solid spec-$z$'s (\texttt{flag\_zspec\_qual} > 1; which require at least two spectral features—either one broad continuum break and one emission line, or two emission lines). For survey design and detailed reductions, please refer to \citet{price2024}. From the ALT DR1 catalog, we use all 1,630 sources with NIRCam/GRISM redshifts over $0.2 < z \leq 8.5$ \citep{naidu2024}; each source has at least one emission line detected with a SNR$>$10. Similarly, we refer readers to \citet{naidu2024} for survey design and detailed reductions. For spectroscopically observed galaxies, we re-fit the photometry with a narrow uniform redshift prior centered at  $z=z_{\text{spec}}\pm0.1$. We then assign this redshift as the ``best'' redshift ($z_{\text{best}}$). The spectroscopic redshifts were first taken from the UNCOVER catalog \citep{price2024}, if available, then from the ALT grism redshift catalog \citep{naidu2024}. If neither is available, then the best-fit photometric redshifts and corresponding posterior distributions were adopted.

\subsection{Sample Selection}\label{subsec:selection}

The main goal of this paper is to compare the SFHs of galaxies in overdense environments and the coeval field. We search for overdensities in redshift space to identify proto-cluster candidates. We select an initial sample of galaxies around cosmic noon with \texttt{USE\_PHOT=1} flag (see \citealt{weaver2024a}), \lgmstar$>$8.5, $1.8< z <3.2$, and filter coverage within the footprint of the MegaScience survey (see \citealt{suess2024} for a detailed footprint). All galaxies in this mass-selected sample have $m_{\text{F444W}} < 26.5$ AB, making them significantly brighter than the F444W 5$\sigma$ point-source detection limit \cite[28.24 AB; ][]{weaver2024a} and indicating that our sample is well within the mass-complete regime. This initial sample ensures that the galaxies’ SEDs are sampled by all broad- and medium-bands over the deepest region of observation, resulting in the most accurate photo-$z$'s.

To quantify the accuracy of our photometric redshifts in this regime, we compare the photo-$z$'s with the spec-$z$'s of sources. In the right panel of Figure~\ref{fig:z_hist}, we present the performance from our \pros photo-$z$'s against the UNCOVER and ALT spec-$z$'s in the range $1.8 < z < 3.2$. We find a NMAD of $\sigma_{\text{NMAD}} \approx0.010$ in this redshift slice, which is comparable to the NMAD found by \citet{suess2024} ($\sigma_{\text{NMAD}}\approx0.015$) over the entire sample. We note that the extreme photo-$z$ outliers (e.g., $z_{\text{phot}} \sim 0.5$ to $z_{\text{spec}} \sim 2.6$) are examples of featureless high-$z$ continua misidentified as low-$z$ quiescent galaxies. See \citet{cutler2025} for further detail on identifying low-$z$ cluster quiescent galaxies.   

\begin{figure}
    \centering
    \includegraphics[width=0.98\linewidth]{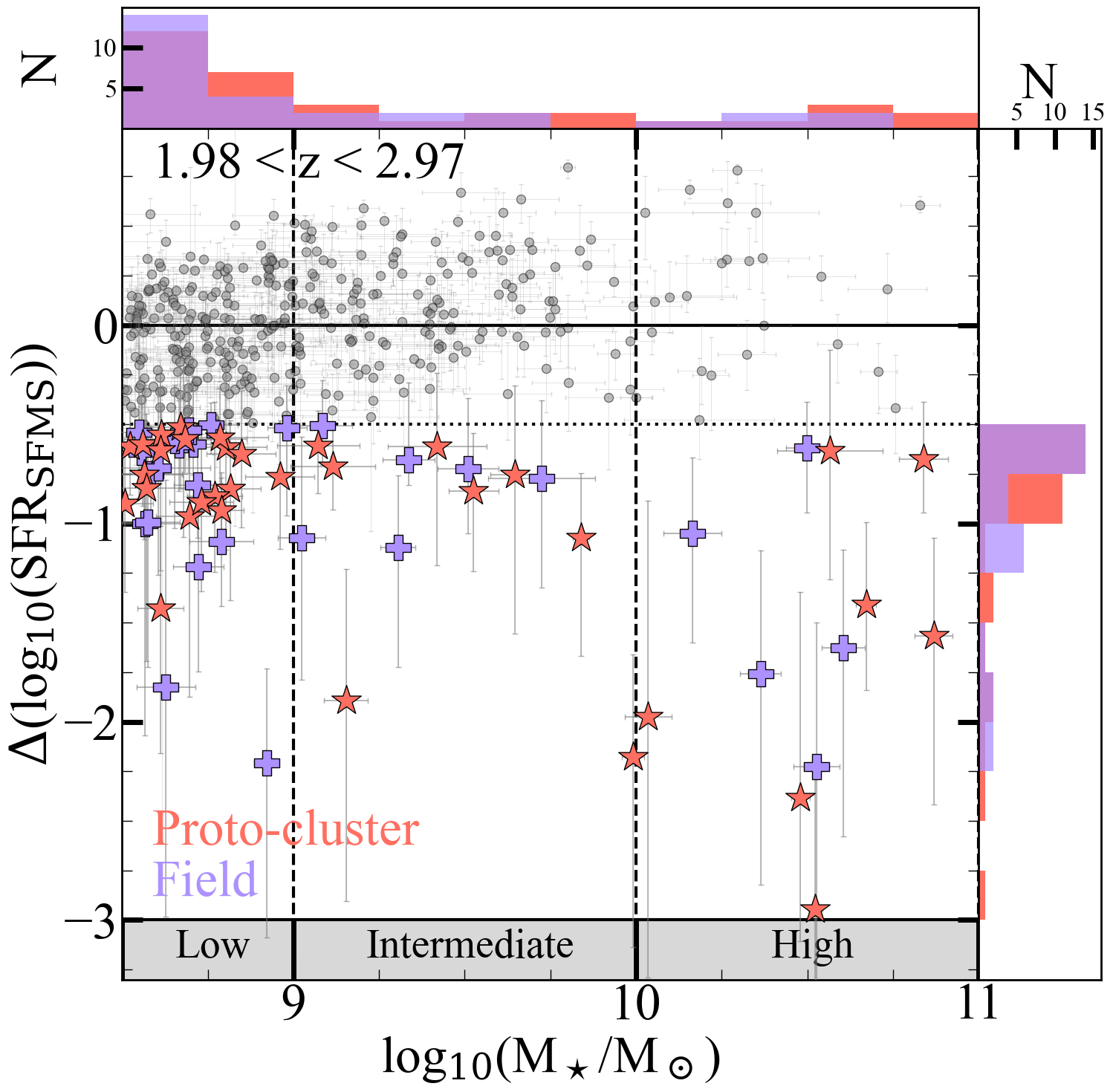}
    \caption{We select quiescent galaxies by the relative difference, $\Delta\log_{10}(\text{SFR}_{\text{SFMS}})$, between their $\log_{10}(\text{SFR}_{100})$ and the SFMS at their redshift from \citet{leja2022} as a function of stellar mass. Quiescent galaxies are selected to be at least 0.5 dex below (dotted line) the main-sequence of star-forming galaxies from \cite{leja2022} (solid curve). Quiescent galaxies in the proto-cluster and in the field are shown as red stars and purple crosses, respectively. We find a total of 217 proto-cluster members (34 quiescent galaxies) and 216 field members (29 quiescent galaxies). As shown in the histograms in the top and right panels, we find that the two subsamples are distributed similarly.}
    \label{fig:deltasfr_mass}
\end{figure}

We plot the redshift distribution of galaxies from this initial sample (top left panel in Figure~\ref{fig:z_hist}) and identify an overdensity peak at $z = 2.58 \pm{0.07}$. We define the uncertainty of our redshift peak, $\sigma= \pm{0.07}$, as approximately two times $\Delta z = 0.036$, where $\Delta z = \sigma_{\text{NMAD}} \times (1+z)$. This yields an overdensity range between $2.51 < z < 2.65$ totaling 217 galaxies. We show the photo-$z$ distribution of spectroscopically observed galaxies in the bottom left panel in Figure~\ref{fig:z_hist} and find that the distribution is well-centered on the redshift peak. We note that this redshift overdensity is also detected in the ALT survey from HeI and Pa$\gamma$ emission at $z\approx2.56$, fully consistent with our photometric findings \citep{naidu2024}.

By summing the observed stellar masses of galaxies in this overdensity and then using the stellar mass-halo mass relation at $z\sim2.5$ \citep{behroozi2019}, we estimate that this overdensity resides in a dark matter halo with a mass of $\sim10^{13.2}\text{M}_{\odot}$ if we include only spectroscopically-confirmed galaxies, and $\sim10^{13.6}\text{M}_{\odot}$ if we also include photometric candidates. These estimates are consistent with the inferred halo masses of confirmed proto-clusters at similar redshifts, including the Spiderweb proto-cluster \citep[$10^{13.5}\text{M}_{\odot}$,][]{dimascolo2023}, a proto-cluster core within the Hyperion structure \citep[$10^{13.5}\text{M}_{\odot}$,][]{champagne2021}, the ODIN proto-clusters \citep[$10^{13.4}\text{M}_{\odot}$,][]{ramakrishnan2025}, and CLJ1001 \citep[$10^{13.7}\text{M}_{\odot}$,][]{wang2016}. Using simulation-based predictions that trace the growth of present-day clusters from their $z\approx 2$ progenitors, we infer that our $z\sim2.6$ overdensity would likely evolve to a ``Coma''-type $z\sim0$ cluster with a total mass of $\sim10^{14.1-14.8}\text{M}_{\odot}$ \citep{muldrew2015}. Given that the projected spatial extent of our system (r$\sim$3.97 cMpc) is smaller than the typical proto-cluster radius at $z \sim 2.5$ from simulations (>6 cMpc), it is likely that we are observing the central group within a larger proto-cluster, and that the total halo mass of the full structure may be even higher \citep{muldrew2015,chiang2017}. We therefore refer to this overdensity as a proto-cluster hereafter. 

Our field comparison sample is selected from redshift ranges close to the proto-cluster at $z\approx2.58$. We avoid $z\approx 2.4$ and $z\approx3.1$, which from the left panel of Figure~\ref{fig:z_hist} may also represent slightly overdense environments. Given this constraint, we select two redshift windows above and below the proto-cluster that yields a similar number of members. This leads to two redshift ranges centered at $z = 2.13 \pm 0.15$ and $z = 2.84 \pm 0.13$ roughly corresponding to $\pm{400}$ Myr of the proto-cluster and yielding 216 field galaxies. Averaging over a higher- and lower-redshift bin helps mitigate any potential environmental effects from the age of the universe. Our conclusions in this work do not vary with small redshift range changes for the proto-cluster and field.

Next, we select quiescent galaxies as those with star-formation rates (SFRs) averaged over the past 100~Myr that lie more than 0.5 dex below the star-forming main sequence (SFMS) at the corresponding epoch, following \citet{leja2022}. We note that at $z\approx2.7$ the SFMS from \citet{leja2022} only extends down to \lgmstar$>$10.2. Therefore we extrapolate this selection to our lower mass limit (\lgmstar$\sim$8.5). This extrapolation is well-motivated by previous studies which have found that the slope of the main sequence is constant down to lower masses, likely because the physics of star formation does not dramatically differ below \lgmstar$<$10.2 \citep{whitaker2014, leja2022, merida2023}. An alternative method for selecting low-mass quiescent galaxies is to use a rest-frame $U-V$ vs $V-J$ color-color selection ($UVJ$ diagram) without the $U-V$ lower limit cut \citep{belli2019, alberts2024, cutler2024}. Such a cut would allow for the selection of younger low-mass quiescent galaxies coming out of a star-burst phase with bluer U-V colors \citep{alberts2024}. We test our selection and arrive at similar conclusions regardless of whether we use an SFR or a $UVJ$ cut to select quiescent galaxies; see Appendix \ref{app:uvj} for further details. 

Figure~\ref{fig:deltasfr_mass} shows our full sample of galaxies (217 in the proto-cluster, 216 in the field) as well as our quiescent subsample (34 in the proto-cluster, and 29 in the field). Marginal histograms show the distributions of our quiescent proto-cluster and field sample; the two samples have similar stellar mass and SFR distributions.

Throughout the rest of this Letter, we split our quiescent sample into three regimes of stellar mass: low-mass (8.5$<$log(M$_{\star}$/M$_{\odot})$$\leq$9.0), intermediate-mass (9.0$<$log(M$_{\star}$/M$_{\odot})$$\leq$10.0), and high-mass (10.0$<$log(M$_{\star}$/M$_{\odot})$$\leq$11.0). We find N = 19 (18) low-mass, 8 (6) intermediate-mass, and 7 (5) high-mass proto-cluster (field) quiescent galaxies. In total, our quiescent sample includes 34 galaxies in the proto-cluster and 29 galaxies in the field. 
\begin{figure*}
    \centering
    \includegraphics[width=\linewidth]{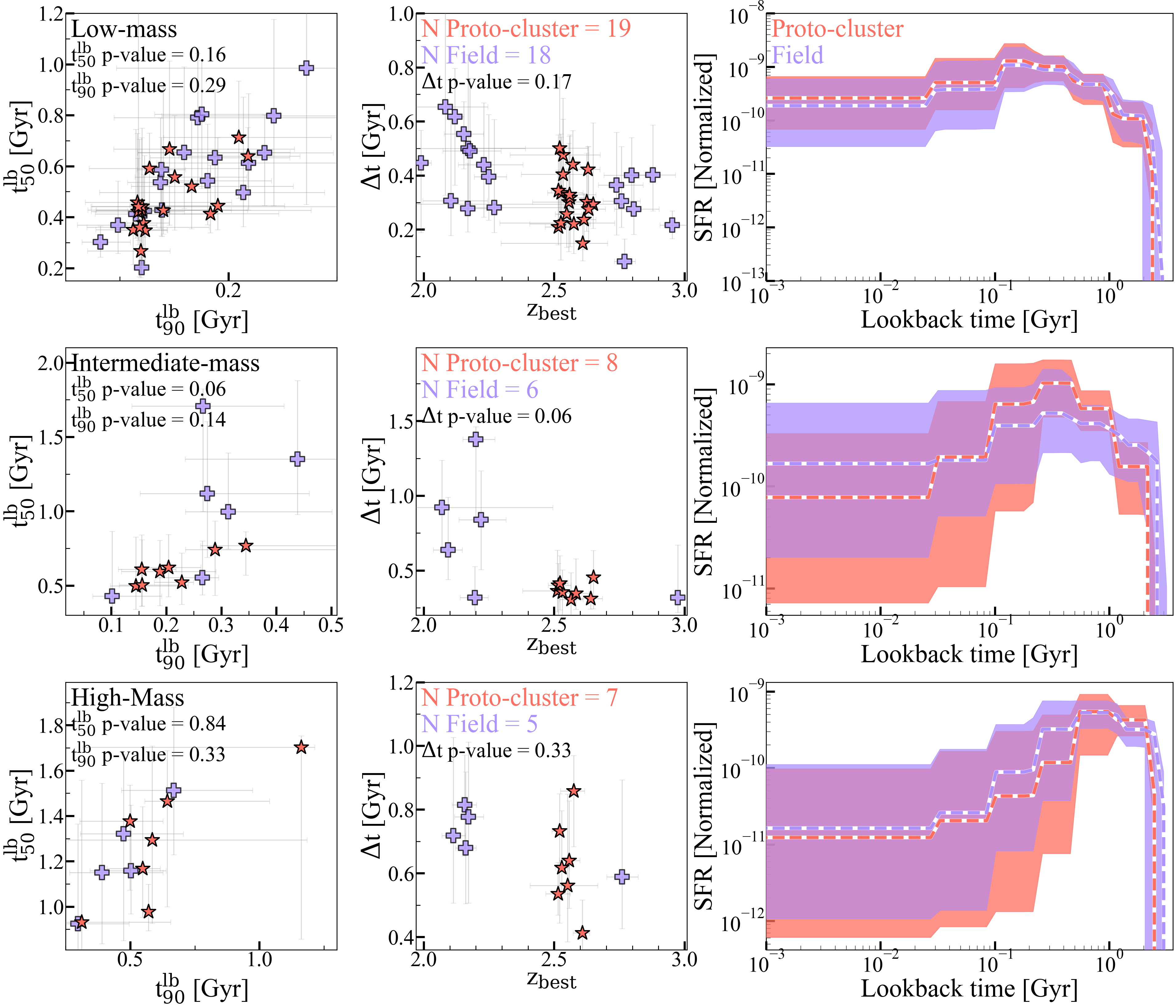}
    \caption{Left: Comparison of the formation time (t$^{\text{lb}}_{50}$) and the quenching time (t$^{\text{lb}}_{90}$) in three stellar mass bins: low-mass (8.5$<$\lgmstar$\leq$9.0), intermediate-mass (9.0$<$\lgmstar$\leq$10.0), and high-mass (10.0$<$\lgmstar$\leq$11.0). Middle: Comparison of  the $\Delta t$ timescale versus the redshift in the same stellar mass bins. Right: Stacked SFHs of the proto-cluster and field galaxies in the same stellar mass bins. The shaded regions indicate the 1$\sigma$ uncertainty while the dashed lines indicate the median SFHs. We compare the distributions in the left and middle panels using a KS test and find them to be statistically indistinguishable. Similarly, the stacked SFHs in the right panels are consistent within 1$\sigma$ demonstrating similar quenching pathways irrespective of environment.} 
    \label{fig:qscales_sfh}
\end{figure*}
\subsection{SFH Timescales}
To quantify potential differences in the stellar populations of proto-cluster members and field members, we calculate the formation and quenching times of our quiescent galaxies from their \pros SFHs, which were derived from SED fitting to the 20-band photometry. We define the formation time as the lookback time when a galaxy formed 50\% of its total mass, $t^{\text{lb}}_{50}$. We consider the quenching time as the lookback time when a galaxy formed 90\% of its total mass, $t^{\text{lb}}_{90}$. Lastly, we establish the star-formation timescale to be the difference of the two ($\Delta t$ = $t^{\text{lb}}_{50}$ - $t^{\text{lb}}_{90}$), roughly indicating how long it took to form the bulk of its recent stellar mass. We calculate $t^{\text{lb}}_{50}$, $t^{\text{lb}}_{90}$, and $\Delta t$ from each draw in the \pros fit and report the 50th percentile of the resulting posterior as the median value and the 16-84th percentile as the 1$\sigma$ error bar. Our focus is to compare these timescales relative to each other (i.e., proto-cluster versus field). We refrain from drawing comparisons to other observational studies that made different prior assumptions on the SFHs given the large impact certain prior choices can have on these derived quantities \citep{leja2019, suess2022}.   

\section{Analysis}\label{sec:analysis}

If the proto-cluster environment at cosmic noon is affecting the quenching mechanisms of galaxies then we might see differences in the SFHs of quiescent galaxies in the proto-cluster compared to the field, as seen in lower-redshift clusters \citep{trager2000,vandokkum2003, thomas2005, renzini2006,thomas2010,muzzin2014, paulinoafonso2020}. We present the comparisons of the SFH timescales and SFHs in Figure~\ref{fig:qscales_sfh}. The left column compares $t^{\text{lb}}_{50}$ and $t^{\text{lb}}_{90}$ of quiescent galaxies in the proto-cluster and field across three stellar mass bins, while the middle column examines the star-formation timescale as a function of redshift. We use a two sample Kolmogorov-Smirnov (KS) test to compare the distributions of the median SFH timescales between environments. The p-values for $t^{\text{lb}}_{50}$ are 0.16, 0.06, and 0.84, for $t^{\text{lb}}_{90}$ are 0.29, 0.14, and 0.33, and for $\Delta t$ are 0.17, 0.06, and 0.33, corresponding to low-, intermediate-, and high-mass bins. All p-values are greater than 0.05, indicating that we cannot reject the hypothesis that the proto-cluster and field samples are drawn from the same parent population. The right column shows the stacked SFHs of galaxies in the proto-cluster versus those in the field. The SFHs are indistinguishable within 1$\sigma$, suggesting that there is no significant difference between quiescent galaxies in the proto-cluster and the field. This finding further supports the results from the KS tests. 

\begin{figure*}
    \centering
    \includegraphics[width=\linewidth]{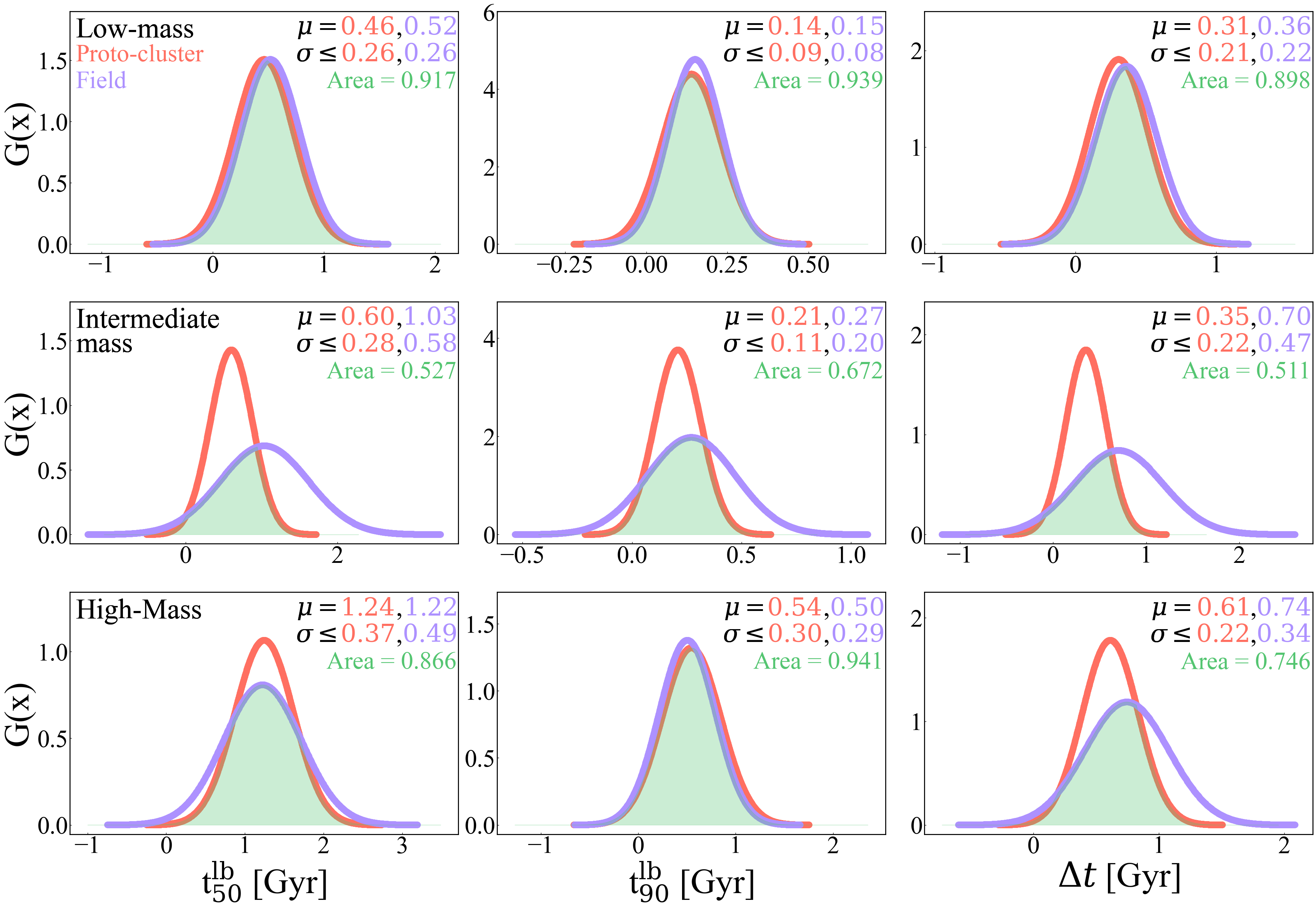}
    \caption{We show the best-fit population models centered at the mean, $\mu$, with a dispersion of $\sigma$. The columns are a comparison of $t^{\text{lb}}_{50}$, $t^{\text{lb}}_{90}$, and $\Delta t$ while the rows are a comparison of the different stellar mass regimes from top to bottom, low-mass, intermediate-mass, and high-mass. The population distribution of the proto-cluster is colored in red while the field is in purple. The overlapping area below the two distributions is in green which roughly quantifies the similarity of the two populations. There is significant overlap between the population models of the proto-cluster and field indicating similar SFH timescales between the two.} 
    \label{fig:population}
\end{figure*}

\subsection{Bayesian Population Modeling}
The KS tests provide a comparison of the median distributions of galaxies in the proto-cluster and the field, but do not account for uncertainties in individual measurements of $t^{\text{lb}}_{50}$, $t^{\text{lb}}_{90}$, and $\Delta t$. To incorporate these uncertainties, we use a Bayesian population modeling framework to infer the underlying population distributions while propagating measurement errors, following the method of \citet{leja2020}. This approach allows us to compare the full posterior distributions of all galaxies collectively, rather than relying solely on median values.

We model the distributions of $t^{\text{lb}}_{50}$, $t^{\text{lb}}_{90}$, and $\Delta t$ for quiescent galaxies as Gaussian functions, leaving the mean and standard deviation of each Gaussian as a free parameter. We separate the data into 3 mass bins (low-, intermediate-, and high-mass), 2 environments (proto-cluster and field), and 3 SFH timescales ($t^{\text{lb}}_{50}$, $t^{\text{lb}}_{90}$, and $\Delta t$) totaling 18 models. The distributions are represented as such:

\begin{equation} \label{eq:gaussian}
    \text{G(x)} = \frac{1}{\sigma \sqrt{2\pi}}e^{-0.5 (\frac{x-\mu_{\text{pop}}}{\sigma_\text{pop}})^{2}}
\end{equation}

where $\mu_\text{pop}$ and $\sigma_\text{pop}$ represent the population mean and dispersion. The observations, \textit{x}, consist of $t^{\text{lb}}_{50}$, $t^{\text{lb}}_{90}$, and $\Delta t$ values for \textit{N} galaxies. In contrast to the KS tests above, which only use a single value for each galaxy (the posterior median), this method incorporates all 1000 posterior draws for each galaxy. The likelihood of observing a single galaxy is computed as the average likelihood over its 1000 draws. The total log-likelihood, $\ln(P)$, for all \textit{N} galaxies is given by:

\begin{equation} \label{eq:popmodel}
    \ln(P) \approx \sum^{N}_{i=1}\ln{ {\Bigg(} \frac{\sum^{M}_{j=1} g(x_{i,j})}{M} {\Bigg)}}
\end{equation}

Before fitting, we set uniform priors on our fit parameters, $\mu_\text{pop}$ and $\sigma_\text{pop}$, of each stellar mass bin and environment. The prior range for $\mu_\text{pop}$ is set by the minimum and maximum observed values, ensuring that the population mean lies within the observed data range. For $\sigma_\text{pop}$, we use a lower bound of $u$, the typical measurement uncertainty in each bin, as values below this are indistinguishable from measurement noise. The upper bound is set to 2 Gyrs, which is sufficient given the age of the universe at our upper limit, $z = 3.0$ ($\sim 2.1$ Gyr). 

After fitting, we perform mock recovery tests to assess the accuracy of our population modeling. We generate \textit{N} mock galaxy observations sampled from a normal distribution with a fixed $\mu_\text{pop}$ and varying $\sigma_\text{pop}$, producing a distribution of observed means, $\mu_\text{obs, n}$ where n is each individual galaxy. For each mock galaxy, we generate 1000 mock draws by sampling a normal distribution with $\mu = \mu_\text{obs, n}$, and $\sigma = u$. The number of galaxies, \textit{N}, is varied between 5 and 20, matching our minimum and maximum sample sizes for various mass and environment bins. Our tests show that  $\mu_\text{pop}$ is accurately recovered within 1$\sigma$ for any input $\sigma_\text{pop}$ and $u$. However, $\sigma_\text{pop}$ is not well recovered for $\sigma_\text{pop} < u$, as the recovered dispersion remains at $u$, indicating that measurement noise and intrinsic dispersion are indistinguishable below this threshold. Thus we set the lower bound of $\sigma_\text{pop}$ to be $u$, preventing the model from inferring an intrinsic dispersion below where it becomes indistinguishable from measurement noise. Lastly, the best-fit recovered $\mu_\text{pop}$ and $\sigma_\text{pop}$ remains unchanged with varying \textit{N} and their uncertainties decrease as the sample size increases. These tests ensure the robustness of our modeling, allowing us to draw reliable conclusions about the population properties of these samples.

Figure~\ref{fig:population} presents the results of our Bayesian population modeling in which we compare the population distributions of the proto-cluster and field across stellar mass bins (rows) and SFH timescales (columns). We report $\sigma_\text{pop}$ as upper limits because the inferred values are consistent with our lower bounds, where it becomes impossible to  distinguish intrinsic dispersion from measurement noise. Our analysis finds that the population models of quiescent galaxies in the proto-cluster and the field are quite similar, with a typical overlapping area $\geq 75\%$. The largest difference occurs in the intermediate-mass bin for $\Delta t$ (area $\sim 50\%$) in which the proto-cluster quiescent galaxies may have slightly shorter star-formation timescales. The $\Delta t$ distributions for intermediate-mass proto-cluster galaxies and field galaxies ($\mu_{\text{pop}} = 0.35,\ \sigma_{\text{pop}} \leq 0.22$ and $\mu_{\text{pop}} = 0.70,\ \sigma_{\text{pop}} \leq 0.47$ respectively) do differ as is also the case for $t^{\text{lb}}_{50}$ (middle left panel in Figure~\ref{fig:population}). However, these differences are within 2$\sigma$ and may not indicate large intrinsic differences in the populations, especially when considering the uneven distribution of field galaxies (1 galaxy in the higher-redshift intermediate mass field bin; central panel in Figure~\ref{fig:qscales_sfh}). 

We verify that our results are not driven by our exact choice of binning scheme. Instead of assigning each galaxy to a bin based on its median properties, we individually bin each posterior draw for each galaxy depending on its stellar mass, SFR, and redshift. For galaxies with large uncertainties or those near bin edges, this approach allows for some of their posterior draws to fall in different bins. We repeat our analysis and find that our results remain unchanged: the differences between the proto-cluster and field populations are still statistically insignificant, with large overlaps between their distributions. Further improving our analysis would require larger samples, a truly coeval field sample, and more precise age measurements to constrain the population distribution.

We also verify that our results are robust to the exact way we measure timescales from the SFH. We measure the quenching timescale, $\Delta t_{\text{quench}}$ \citep[following][]{carnall2018, park2024}, defined as the difference between the lookback time when the SFR peaked ($t^{\text{lb}}_{\text{peak}}$) and the lookback time when the SFR reached 20\% of the peak value ($t^{\text{lb}}_{\text{20\%}}$). Using this metric, we compare the median quenching timescales and inferred population models between the proto-cluster and field samples. Again, this comparison highlights the large KS test p-values ($>$0.05) and a typical overlapping area of approximately 90\%, indicating that there is no significant difference in the field and proto-cluster populations. 

Overall, these results support the KS test results that the SFHs of quiescent galaxies in overdense environments at $z\sim2.6$ are statistically indistinguishable from those in the field across the stellar mass range 8.5$<$\lgmstar$\leq$11.0. 

\begin{figure}
    \centering
    \includegraphics[width=0.98\linewidth]{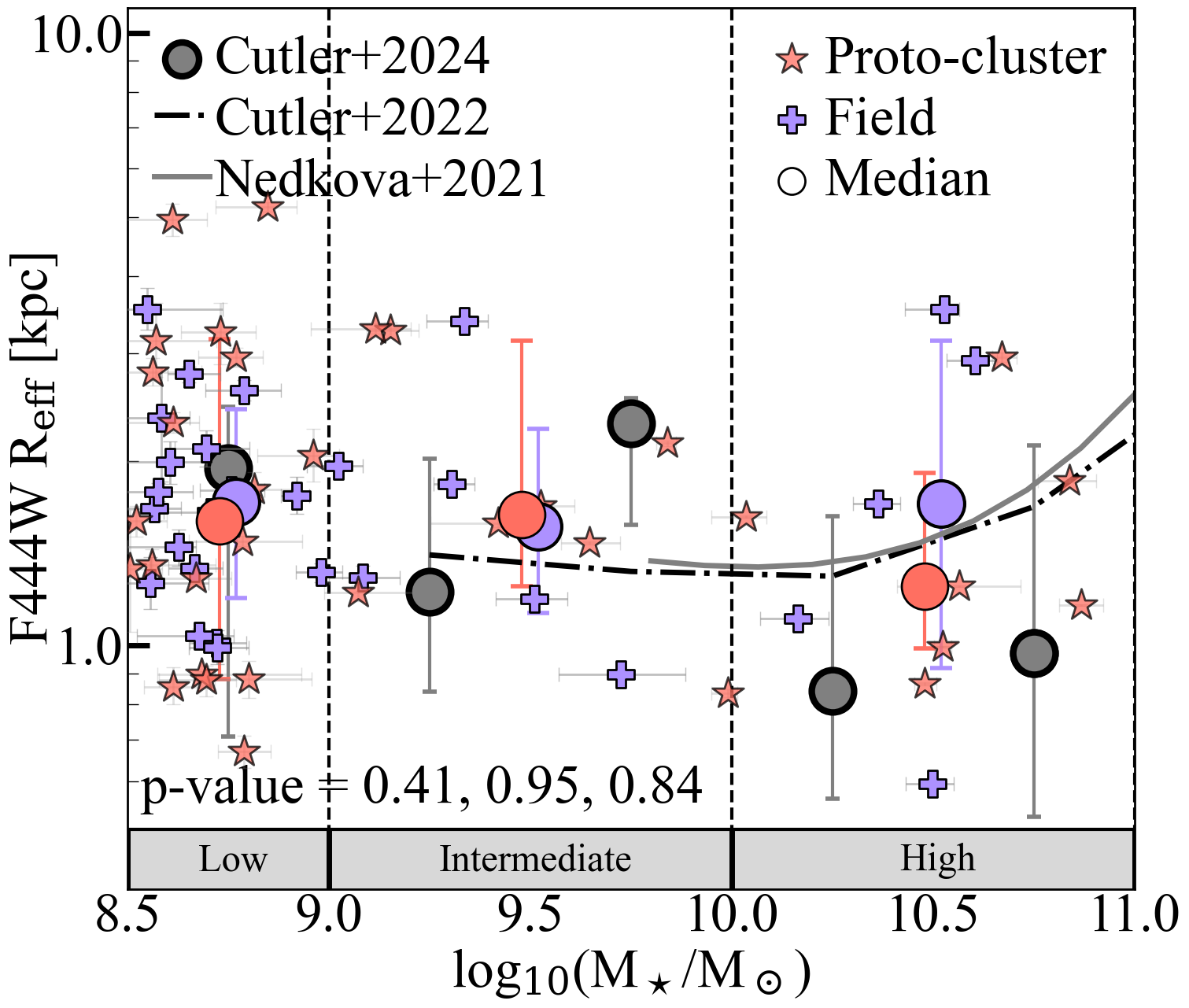}
    \caption{Mass-size relation of quiescent galaxies in the proto-cluster (red stars) and the field (blue crosses), measured in the JWST NIRCam F444W. Their median mass-size are shown as red and blue circles, respectively, along with median values from \citet{cutler2024} (grey circles). We also overplot the mass-size relation from HST F160W measurements from \citet{nedkova2021} (solid black line) and \citet{cutler2022} (dot dash line). We find a similarly flat mass-size relation and no significant difference between the proto-cluster and field populations.}
    \label{fig:mass_size}
\end{figure}

\subsection{Sizes}
We investigate the sizes of the two populations to see whether the overdense environment has affected their structural properties, as seen in previous works \citep{kuchner2017,carlsten2021}. We use \pysersic, a Bayesian inference fitting tool \citep{pashamiller2023}, to fit 1D S\'{e}rsic profiles to the F444W images. In this redshift regime, F444W probes rest-frame $\sim 1.2~\mu m$ where we expect the light to trace the stellar mass distribution of galaxies \citep{bell2001}. The methods are described in detail in \textcolor{blue}{Zhang et al. (in prep.)} and \citet{miller2024}. In brief, we simultaneously fit all other sources within 2 arcsecs of the main object, within 2 magnitudes in F277W, F356W, and F444W of the main objects, and a SNR > 5 in F277W, F356W, and F444W. We mask any other sources in the UNCOVER DR3 segmentation map. We draw samples from the posterior using a No U-turn sampler using 2 chains with 750 warm-up and 500 sampling steps each \citep{hoffman2014, phan2019}.

We show the mass-size relation in Figure~\ref{fig:mass_size} with the circular scatter points illustrating the median effective radii (semi-major axis) in a given stellar mass bin. We perform a KS test in each stellar mass bin comparing the sizes of quiescent galaxies in the proto-cluster and those in the field. Because of the large p-values of 0.41, 0.95, and 0.84 (low-, intermediate-, high-mass), we cannot reject the hypothesis that the proto-cluster and field samples are drawn from the same parent population. We note that our median sizes are similar to those measured by  \citet{nedkova2021} (HST F160W), \citet{cutler2022} (HST F160W) and \citet{cutler2024} (JWST F444W), all of which use the \textsc{GALFIT} fitting tool \citep{peng2002, peng2010a} and measure primarily field galaxies. 

\subsection{Quiescent Fraction and Quiescent Fraction Excess}

\begin{figure*}
    \centering
    \includegraphics[width=\linewidth]{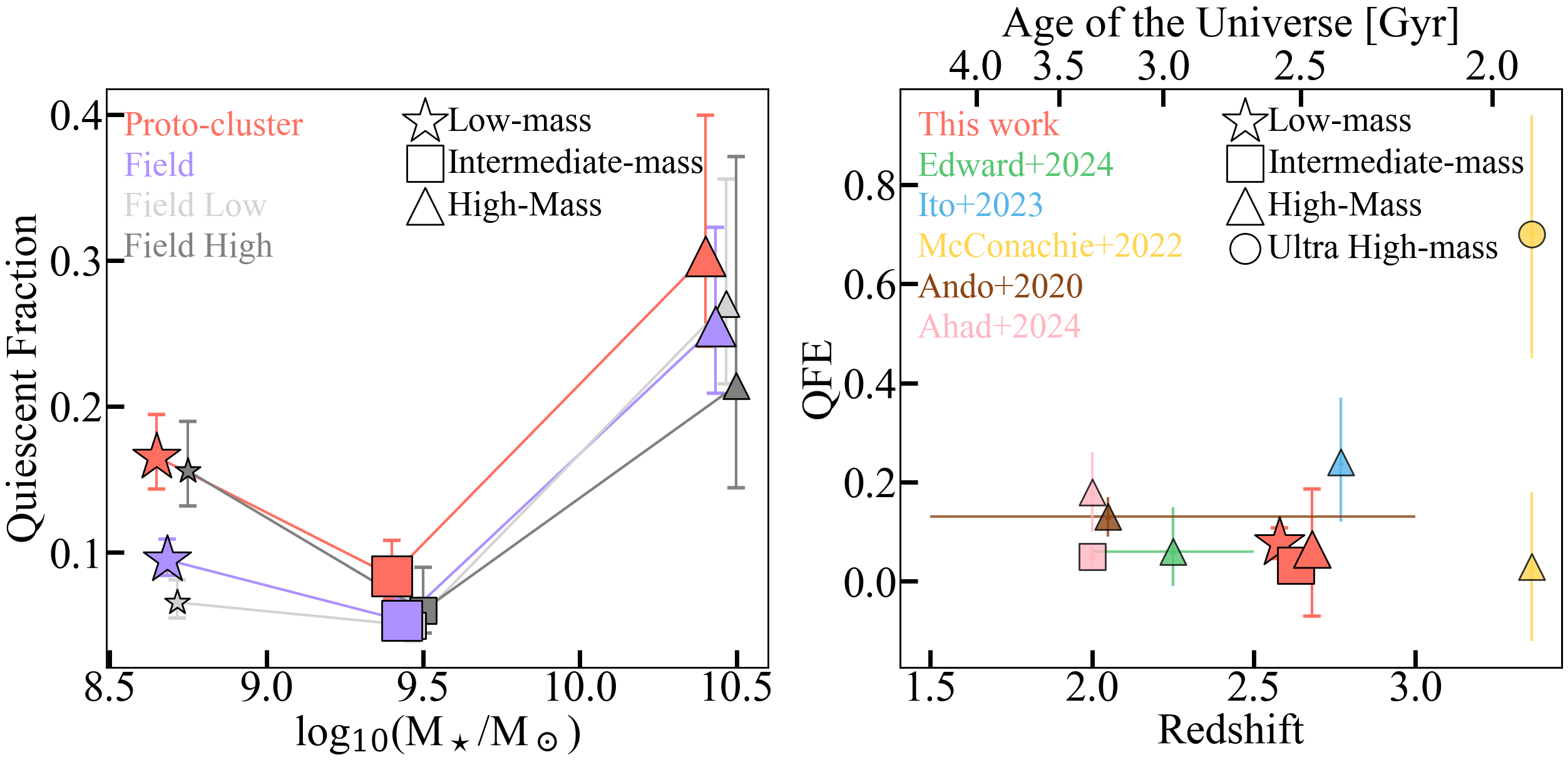}
    \caption{Left: The quiescent fraction for galaxies in the proto-cluster (red), combined field (purple), lower-redshift field (light gray), and higher-redshift field (dark gray). The quenched fractions in the proto-cluster are indistinguishable from the field at high-mass (triangle). The quenched fractions in the intermediate- (square) and low-mass (star) bins are slightly elevated in the proto-cluster, but within $\sim2\sigma$ of those in the field. Right: The QFE for proto-clusters in our work and current literature around cosmic noon. The current literature has primarily investigated the high-mass regime with our work extending down to low-masses. We've differentiated the stellar mass regimes of the QFE by their shapes if applicable. We note that the QFE from \citet{mcconachie2022} (yellow) includes the ultra-high mass galaxies (\lgmstar$>$11.0) and are shown as circles. We also note that half of the sample from \citet{ito2023} (blue) also contains ultra-high mass galaxies and may be driving the elevated QFE. Lastly, the work in \citet{ando2020} (brown) combines several proto-clusters from redshift 1.5 to 3.0 across the intermediate- and high-mass bins. Our analysis is consistent with other studies within $2\sigma$, especially in the similar stellar mass bins, and suggests a relatively low excess in quiescent galaxies.} 
    \label{fig:qfe}
\end{figure*}

Lastly, we measure the quiescent fraction and the quiescent fraction excess (QFE) -- the fraction of galaxies that are quenched in an overdense environment but would still be star-forming in a lower density environment -- to assess whether there is evidence for environmentally driven quenching and to facilitate comparison with previous studies \citep{bahe2017,vanderburg2020, edward2024}. The quiescent fractions, measured down to \lgmstar$\sim$8.5, are shown in the left panel of Figure~\ref{fig:qfe}. This analysis reveals that, in the intermediate- and high-mass regimes, the proto-cluster quiescent fractions are consistent with those of the field within $1\sigma$ uncertainties. At low masses, however, there is a discrepancy between the proto-cluster fraction ($0.17^{+0.03}_{-0.02}$) and the field fraction ($0.10^{+0.01}_{-0.01}$) that exceeds $2\sigma$. We note, however, that this difference is driven only by the lower-redshift field environment: the higher-redshift field environment has a consistent quenched fraction as the proto-cluster. This further underscores the need for a truly coeval sample of field quiescent galaxies in future works covering larger sky areas. 

To compare with previous literature, we measure the QFE following the method of \citet{wetzel2012, bahe2017, vanderburg2020, edward2024} as given by:

\begin{equation} \label{eq:qfe}
    \text{QFE} = \frac{\text{QF}_{\text{pc}}-\text{QF}_{\text{field}}}{1-\text{QF}_{\text{field}}}
\end{equation}
where $\text{QF}_{\text{pc}}$ and $\text{QF}_{\text{field}}$ denote the quiescent fractions in the proto-cluster and the field. To estimate the uncertainties on both the quiescent fraction and QFE, we use the bootstrap method to propagate uncertainties in the number of quiescent and total galaxies and adopt the 16th and 84th percentile of the resulting distributions as the 1$\sigma$ confidence interval. The QFE values are shown on the right panel of Figure~\ref{fig:qfe} and demonstrates that, across the entire stellar mass regime, the proto-cluster exhibits relatively low ($<$0.1) QFE values; $0.08^{+0.02}_{-0.02}$, $0.03^{+0.02}_{-0.01}$, and $0.07^{+0.09}_{-0.07}$ for the low-, intermediate-, and high-mass bins, respectively. This suggests that the overdense environment has not yet developed a substantial excess of quiescent galaxies relative to the field. Our results are consistent within $1\sigma$ with most other studies at cosmic noon, with the notable exception of the ultra high-mass sample (yellow circle) from \citet{mcconachie2022} which may be specific to ultra high-masses. 

\section{Discussion}{\label{sec:discussion}}

In this paper, we examine the SFHs of quiescent galaxies in a proto-cluster at $z\approx2.58$ and find no evidence of environmental quenching in those that reside in an overdense environment as compared to those that reside in the coeval field. Our analysis demonstrates that these subsamples of galaxies have similar formation and quenching times, star-formation timescales, stacked SFHs, and sizes across all stellar mass ranges as shown in Figures~\ref{fig:qscales_sfh}, \ref{fig:population}, and \ref{fig:mass_size}. Similarly, our investigation of the quiescent fraction shows a slight excess of low-mass quiescent galaxies in the proto-cluster compared to the lower-redshift field sample, but the relative excess of quiescent galaxies is small (QFE$<0.1$) and indicates little to no role of environmental quenching. Our results suggest that this proto-cluster environment is not efficiently altering the quenching process in galaxies at $z\sim2.6$.

Our study focuses on comparing the relative ages of low-mass quiescent galaxies in the proto-cluster and the field. The absolute calibration of galaxy ages -- especially mass-weighted ages like those we use here -- depend on the choice of SFH priors (e.g., \citealt{leja2019, suess2022}; \textcolor{blue}{Gallazzi et al. in prep.}). 
Because of this, we refrain from directly comparing our timescales with the timescales observed in other works. However, the agreement in measured ages between the proto-cluster and field samples suggests that similar quenching mechanisms are acting on both populations and that these mechanisms are unlikely to be environmentally driven. Further observations with medium-resolution spectroscopy could break the age-metallicity degeneracy and improve the absolute calibration of our ages, allowing for more precise comparisons to theoretical predictions.

Our results are consistent with some previous ground-based observational studies which have found no evidence for environmental quenching at cosmic noon \citep{edward2024, singh2024, forrest2024}. \citet{edward2024} and \citet{forrest2024} found similar quenched fractions in proto-clusters and the field, particularly in their intermediate-mass bin, and argue that proto-cluster environments are not accelerating quenching. Similarly, we find that the quiescent fractions in our proto-cluster are statistically consistent with those of the field across intermediate- and high-mass bins, and while there is a mild excess at low mass, the corresponding QFE values remain low.

Other studies targeting ultra high-mass galaxies \citep[e.g., ][]{ito2023,mcconachie2024} find a significant excess (QFE $\gtrsim$ 0.2) of quiescent galaxies that may suggest environmental effects are at play. However, recent simulation results from \citet{ahad2024} suggests that this excess could arise from differences in the halo mass distributions rather than environmental quenching. At fixed stellar mass, galaxies residing in more massive halos are more likely to be quenched. In overdense regions, the combination of a top-heavy halo mass function and a normal stellar mass function naturally leads to a higher quenched fraction, independent of any environmental quenching processes \citep{ahad2024}. 

At intermediate-masses, \cite{ahad2024} find that the role of environmental quenching appears to increase between $2.0<z<3.5$ as indicated by an elevated quiescent fraction in overdense environments; however, the relative excess remains low (QFE$<0.1$, Fig.~\ref{fig:qfe}) suggesting a more complicated interpretation. Further observational studies and simulation predictions targeting low-mass galaxies in overdense environments are needed to enable direct comparisons with our findings.

One implication from the results found in \cite{ahad2024} is that there may be a diversity among proto-clusters depending on the intrinsic halo mass function of the galaxy population. Recent studies of highly clustered, high-mass galaxies in the CLJ1001 and Hyperion proto-clusters at $z\approx2.5$ have identified potential signs of environmental quenching, including ram-pressure stripping \citep{xu2025} and rapidly depleted cold molecular gas reservoirs \citep{gururajan2025}. Similarly, in the Spiderweb proto-cluster at $z\approx2.16$, observations have revealed a developing proto-ICM \citep{dimascolo2023, lepore2024}, a likely precursor to environmentally driven quenching. In contrast, other studies of filaments within the Hyperion proto-cluster report no evidence for a well-developed ICM capable of stripping infalling galaxies \citep{champagne2021} and our work indicates no evidence of an ICM based on the similar SFHs and sizes of proto-cluster and field quiescent galaxies. These contrasting findings may be explained by differences in the underlying dark matter halo mass functions. For example, a bottom-heavy distribution in CLJ1001 may result in shallower potential wells, limiting their ability to retain cold gas and making galaxies more susceptible to environmental effects. These findings highlight that proto-clusters are not a uniform population, and that the presence or absence of environmental quenching may depend sensitively on the underlying halo mass distribution and evolutionary state of each structure.

Observational constraints on these environments remain limited by potential uncertainties in identifying and characterizing proto-cluster membership, particularly at high redshift. Associating galaxies with a physical proto-cluster using photometric redshifts will always be complicated given uncertainties from SED modeling. However, we mitigate much of this uncertainty by including deep medium band photometry, which we demonstrate provide the necessary average precision for this task ($\sigma_{\text{NMAD}}\sim~0.01*(1+z)$, Fig.~\ref{fig:z_hist}). Future spectroscopic surveys will be critical to ensure complete selection of galaxies in the proto-cluster, constrain their stellar populations and star-formation histories, and confirm their quiescence. Even still, small extragalactic fields will have limitations with cosmic variance (i.e., unable to observe a variety of proto-clusters) and constructing truly coeval field samples (i.e., at the same exact redshift), underscoring the need for similar quality data in other fields. Much larger fields are on the horizon (e.g., Technicolor, GO-3362, PI: Muzzin; JUMPS, GO-5890, PI: Withers; MINERVA, GO-7814 , PIs: Marchesini, Muzzin, \& Suess; SPAM, GO-8559, PI: Davis \& Larson) and will provide more stringent and robust tests on the emergence of environmental quenching on the proto-cluster galaxy population at cosmic noon. Further morphological analysis to determine \textit{where} quenching occurs within galaxies, such as inside-out versus outside-in quenching, is also crucial for gaining a deeper understanding of the mechanisms driving quenching in low-mass galaxies \citep{bezanson2009, suess2019, albertsnoble2022}. 

By extending quenching measurements into a stellar mass regime (\lgmstar$\sim$8.5 at $z\sim2-3$) previously inaccessible to HST or ground-based facilities, this work furthers our understanding of proto-cluster environments at cosmic noon and provides new insight into the onset of environmental quenching in low-mass galaxies. Despite revealing this unique population, our results align with previous observational studies: proto-cluster environments at $z\sim 2-3$ do not enhance or accelerate quenching compared to the field. This indicates that environmental quenching is not yet the primary driver in suppressing the star formation of these galaxies at these redshifts, likely due to the lack of a sufficiently dense and hot ICM to effectively strip cold gas. However, it is clear from low-redshift studies that environmental quenching must eventually occur \citep{Dressler1980, Balogh2004, Kauffmann2004, Baldry2006, peng2010b, muzzin2014,hamadouche2024,cutler2025} and future work will \textit{uncover} when this happens.

\section{Acknowledgments}
This work is based in part on observations made with the NASA/ESA/CSA \textit{James Webb Space Telescope}. The data were obtained from the Mikulski Archive for Space Telescopes at the Space Telescope Science Institute, which is operated by the Association of Universities for Research in Astronomy, Inc., under NASA contract NAS 5-03127 for the JWST. These observations are associated with the JWST Cycle 1 GO program \#2561 and Cycle 2 GO program \#4111. The JWST data presented in this article were obtained from the Mikulski Archive for Space Telescopes (MAST) at the Space Telescope Science Institute. Support for program JWST-GO-2561 and JWST-GO-4111 was provided by NASA through a grant from the Space Telescope Science Institute, which is operated by the Associations of Universities for Research in Astronomy, Incorporated, under NASA contract NAS5-26555.
This research was supported in part by the University of Pittsburgh Center for Research Computing, RRID:SCR\_022735, through the resources provided. Specifically, this work used the H2P cluster, which is supported by NSF award number OAC-2117681.
Cloud-based data processing and file storage for this work is provided by the AWS Cloud Credits for Research program. 
The Cosmic Dawn Center is funded by the Danish National Research Foundation (DNRF) under grant \#140.
The BGU lensing group acknowledges support by grant No.~2020750 from the United States-Israel Binational Science Foundation (BSF) and grant No.~2109066 from the United States National Science Foundation (NSF), by the Israel Science Foundation Grant No.~864/23, and by the Ministry of Science \& Technology, Israel.
Computations for this research were performed on the Pennsylvania State University's Institute for Computational and Data Sciences' Roar supercomputer.

\facilities{JWST (NIRSpec, NIRCam), HST (ACS, WFC3)}

\software{
    \textsc{astropy} \citep{astropy2022}, 
    \textsc{msaexp} (v0.8.5; \citep{msaexp2022}), 
    \texttt{jwst} pipeline (v1.14.0; \citep{jwst2022}), 
    \grizli \citep[\url{github.com/gbrammer/grizli}]{brammer2023}, 
    \eazy \citep{brammer2008}, 
    \textsc{matplotlib} \citep{matplotlib2007},  
    \textsc{numpy} \citep{numpy2011}, 
    \textsc{Dynesty} \citep{speagle2020},
    \pros \citep{johnson2021}
    }
\appendix
\begin{figure*}
    \centering
    \includegraphics[width=\linewidth]{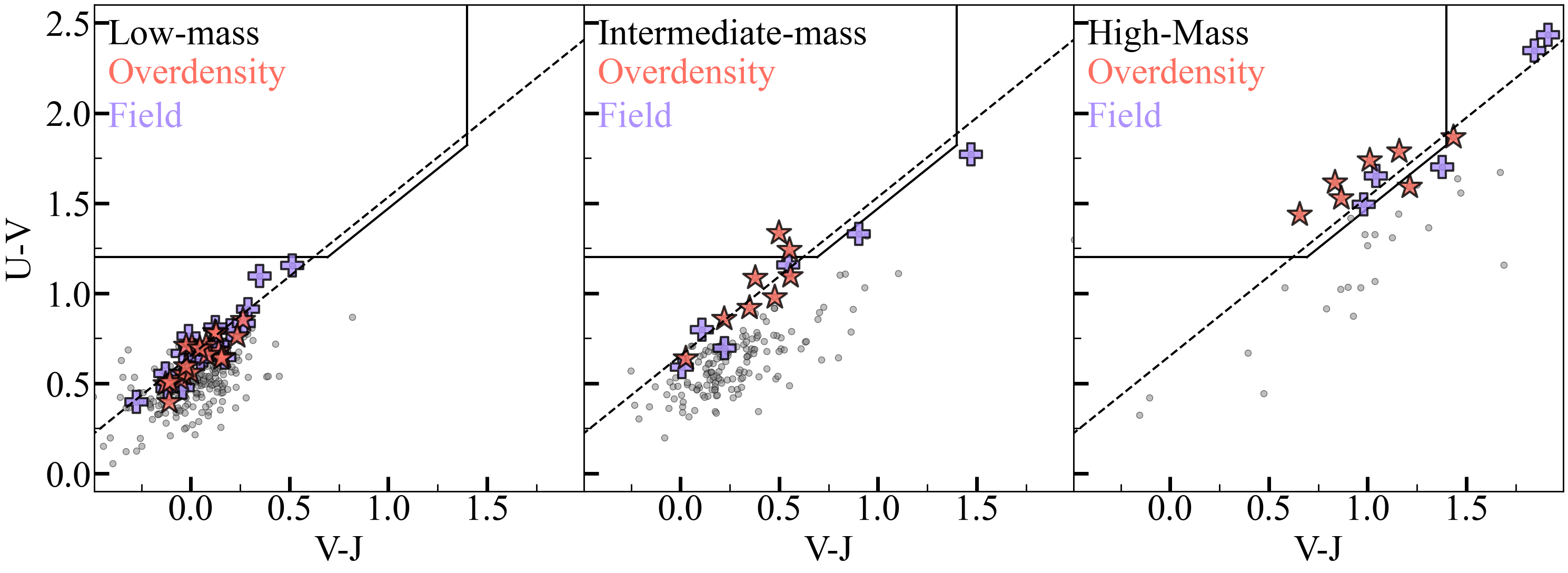}
    \caption{$UVJ$ color-color diagram in the three stellar mass bins. Similarly to Fig.~\ref{fig:deltasfr_mass}, red stars indicate quiescent galaxies in the proto-cluster, purple crosses indicate quiescent galaxies in the field, and the grey points indicate all other galaxies in the proto-cluster or field. The solid lines indicate the traditional quiescent wedge from \cite{whitaker2011} and the solid dashed line indicates the \cite{belli2019} selection. We find that all low-mass galaxies would otherwise be not selected using the $UVJ$ diagram. There is significant overlap between quiescent galaxies in the proto-cluster and field suggesting that they have similar stellar population properties.} 
    \label{fig:uvj}
\end{figure*}

\section{Light-Weighted Ages}\label{app:uvj}
To further verify our key result that the ages of proto-cluster and field galaxies are consistent, we show in Figure \ref{fig:uvj} the positions of our galaxies on the $UVJ$ color-color diagram. Previous works \citep[e.g.,][]{whitaker_2012_psb, belli2019} suggest that the location of quiescent galaxies in this color-color space is a good indicator of their age with younger quiescent galaxies typically exhibiting bluer $U-V$ and $V-J$ colors, and older ones appearing redder (though metallicity and dust attenuation effects are also important, see \citealt{cheng2025}). Figure~\ref{fig:uvj} shows that the low-mass quiescent galaxies in the proto-cluster have comparable $UVJ$ colors as the low-mass quiescent galaxies in the field suggesting similar light-weighted ages. We find that these low-mass quiescent galaxies occupy a similar region of the $UVJ$ diagram as the relatively young low-mass quiescent galaxies in \cite{cutler2024}, but note that there are potential differences when comparing light-weighted and mass-weighted ages. The similar locations on the $UVJ$ diagram show that comparing stellar ages of quiescent galaxies between the two environments using either light-weighted or mass-weighted ages suggests no significant differences.

\bibliography{bibliography}{}
\bibliographystyle{aasjournal}


\end{CJK*}
\end{document}